\documentclass[prd,aps,preprint,amsmath,amssymb,eshowkeys,nofootinbib]{revtex4-1}
\usepackage[mathscr]{euscript}
\usepackage{dcolumn}
\usepackage{bm}
\usepackage{graphicx}
\usepackage{subfigure}
\usepackage{slashbox}
\usepackage{amsmath,amssymb,amsthm}
\usepackage[colorlinks=true,linkcolor=red,urlcolor=green,citecolor=green]{hyperref}
\newcommand{\bea}{\begin{eqnarray}}
\newcommand{\eea}{\end{eqnarray}}
\newcommand{\beq}{\begin{equation}}
\newcommand{\eeq}{\end{equation}}

\def\/{\over}

\begin{document}

\title{Holographic inflation and holographic dark energy from entropy of the anti-de Sitter black hole}
\author{Qihong Huang$^{1}$\footnote{Corresponding author: huangqihongzynu@163.com}, He Huang$^{2}$, Bing Xu$^{3}$ and Kaituo Zhang$^{4}$}
\affiliation{
$^1$ School of Physics and Electronic Science, Zunyi Normal University, Zunyi, Guizhou 563006, China\\
$^2$ Institute of Applied Mechanics, Zhejiang University, Hangzhou, Zhejiang 310058, China\\
$^3$ School of Electrical and Electronic Engineering, Anhui Science and Technology University, Bengbu, Anhui 233030, China\\
$^4$ Department of Physics, Anhui Normal University, Wuhu, Anhui 241000, China
}

\begin{abstract}
Based on the entropy of anti-de Sitter black hole, a new holographic dark energy model has been proposed. When the Hubble horizon and particle horizon are chosen as the IR cutoff, the late-time accelerated expansion of universe is realized. In this paper, we consider the Hubble horizon as the IR cutoff to investigate holographic inflation and slow-roll inflation in this model. We find that slow-roll inflation with the chaotic potential $V_{0}\phi^{n}$ is favored by Planck results for some special cases, such as $n=1/3$ and $n=1/2$, while holographic inflation is not supported by Planck results. Then, we analyze the reheating temperature and the number of reheating e-folds in this model, and we find that the results favor the cases $n=1/3$ and $n=1/2$. Finally, we use the dynamical analysis method, statefinder diagnostic pairs, and the Hubble diagram to analyze this model. Our results indicate that when $b^{2}$ takes a small value, this model cannot be distinguished from the standard $\Lambda$CDM model and can serve as an alternative to it.
\end{abstract}

\maketitle

\section{Introduction}

Inflation, an exponentially accelerating expansion in the early evolution of the universe, offers a successful solution to the horizon and flatness problems encountered in standard big bang cosmology~\cite{Guth1981, Linde1982}. Inflation generates adiabatic, Gaussian, and nearly scale-invariant scalar perturbations, which are responsible for the Cosmic Microwave Background (CMB) temperature anisotropies and Large-Scale Structure formation~\cite{Mukhanov1981, Lewis2000, Bernardeau2002}. These perturbations are supported by observational results from COBE, WMAP, and Planck~\cite{Smoot1992, Hinshaw2013, Planck2020}. The simplest inflation model, known as slow-roll inflation~\cite{Linde1982, Noh2001, Weinberg2008}, is driven by a canonical scalar field that slowly rolls down its potential, resulting in exponential expansion. Inflation ends when the potential reaches its minimum value, after which the universe enters a radiation-dominated epoch through a reheating phase~\cite{Shtanov1995, Bassett2006}, during which particles are generated via processes such as leptogenesis, baryogenesis, and nucleosynthesis~\cite{Davidson2008}. After inflation, the information about scalar perturbations is encoded in the primordial scalar power spectrum, described by the scalar spectral index $n_{s}$~\cite{Planck2020a}, while tensor perturbations produced during inflation manifest as primordial gravitational waves, characterized by the tensor-to-scalar ratio $r$~\cite{Planck2020a, Maggiore2018}. Using the measurable scalar spectral index $n_{s}$ and tensor-to-scalar ratio $r$, slow-roll inflation models are tightly constrained by current observational data~\cite{Planck2020a}. Under these constraints, numerous slow-roll inflation models~\cite{Ding2024, Ragavendra2024, Pozdeeva2024, Zhang2024, Marco2024, Lambiase2023, Afshar2023, Bhat2023, Dioguardi2022, Karciauskas2022, Chen2022, Capozziello2021, Forconi2021, Cai2021, Gamonal2021, Fu2020, Akin2020, Fu2019, Granda2019, Gonzalez-Espinoza2019, Granda2019a, Yi2018, Casadio2018, Odintsov2018, Tahmasebzadeh2016, Yang2015, Koh2014, Gao2014, Antusch2014, Guo2010, Satoh2010, Kaneda2010, Brax2009, Tzirakis2009, Barenboim2007, Peiris2006, Gong1999, Barrow1995} and potentials~\cite{Choudhury2024, Choi2022, Fei2020, Barrow2018, Gao2018, Koh2017, Fei2017, Lin2016, Cline2006, Easther2003, Adams1995, Liddle1994} have been constructed. Besides slow-roll inflation, other models such as constant-roll inflation~\cite{Motohashi2015, Gao2017, Gao2018a, Yi2018a, Gao2018b, Motohashi2019, Gao2019, Ravanpak2022, Mohammadi2022, Shokri2022, Panda2023, Liu2024}, curved inflation~\cite{Thavanesan2021, Shumaylov2022, Huang2022, Huang2023, Huang2023a}, and slow expansion~\cite{Liu2011, Liu2013, Cai2016, Huang2019} can also generate primordial power spectra that are nearly scale-invariant and consistent with observations. Recently, a new inflationary model called holographic inflation~\cite{Nojiri2019, Oliveros2019} was proposed, where inflation is driven by holographic dark energy. This model also generates nearly scale-invariant primordial power spectra and is supported by observations. Initially, the particle or future event horizon was considered as the IR cutoff~\cite{Nojiri2019}. Later, with the GO length scale as the IR cutoff, it was found that holographic inflation can be realized in both holographic dark energy~\cite{Oliveros2019} and Tsallis holographic dark energy~\cite{Mohammadi2021}.

After the end of inflation, the universe enters a reheating phase, during which the inflaton field oscillates around the minimum of its potential and gradually decays into other particles~\cite{Bassett2006, Cai2015}. In this phase, radiation and matter are produced through the decay of the inflaton field or other fields. These particles interact and eventually thermalize, reaching equilibrium at a reheating temperature $T_{reh}$. During this phase, the universe must thermalize at an extremely high temperature to facilitate big bang nucleosynthesis and baryogenesis. As a result, the temperature of the universe rises, leading to the radiation-dominated era and ultimately transitioning into the standard hot big bang model. To characterize the reheating phase, several important reheating parameters have been proposed: the reheating temperature $T_{reh}$, which is constrained by big bang nucleosynthesis~\cite{Steigman2007} and the energy scale of inflation, typically lies in the range $10^{-2}GeV \leq T_{reh} \leq 10^{16}GeV$; the effective equation of state parameter $\omega_{reh}$, often assumed to be constant, is bounded within the range $-\frac{1}{3} \leq \omega_{reh} \leq 1$~\cite{Dai2014, Cook2015, Munoz2015, Goswami2018, Zhou2022, Yadav2024, Zhang2021, Zhang2023, Martin2015}; the number of reheating e-folds $N_{reh}$, defined as the number of e-folds from the end of inflation to the beginning of the radiation-dominated era, is influenced by the effective equation of state parameter $\omega_{reh}$ and the inflationary model~\cite{Dai2014, Cook2015, Munoz2015, Goswami2018, Zhou2022, Yadav2024, Zhang2021, Zhang2023}. To describe the reheating phase in the early universe, several reheating models have been proposed, including the perturbative decay of an oscillating inflaton field at the end of inflation~\cite{Dolgov1982, Abbott1982}, as well as non-perturbative processes such as parametric resonance decay or tachyonic instability~\cite{Traschen1990, Shtanov1995, Bassett2006, Fu2017, Fu2019a, Jin2020, Li2020}.

Another acceleration phase is the current accelerated expansion of the universe, which has been confirmed by observations~\cite{Perlmutter1999, Riess1998, Spergel2003, Spergel2007, Tegmark2004, Eisenstein2005}. To explain this late-time acceleration, dark energy has been proposed. One of the leading candidates is holographic dark energy (HDE)~\cite{Hsu2004, Horvat2004, Li2004}, based on the holographic principle, which states that the entropy of a system scales with its surface area~\cite{Witten1998, Bousso2002}. A viable dark energy model must not only explain the current accelerated expansion but also describe the entire evolutionary history of the universe. Specifically, the universe must originate from a radiation-dominated epoch, transition into a matter-dominated epoch to allow large-scale structure formation, and finally evolve into a dark energy-dominated epoch. To qualitatively explore the evolution of the universe in a given cosmological model, the dynamical analysis method is employed. In this method, the behavior of critical points in the dynamical system describes the primary evolutionary stages of the universe, with a stable critical point corresponding to the dark energy-dominated epoch~\cite{Bahamonde2018}. When applied to HDE models, this method has achieved significant success~\cite{Setare2009, Liu2010, Banerjee2015, Mahata2015, Mishra2019, Bargach2019, Tita2024}. Subsequently, it has been extended to other holographic dark energy models, such as Tsallis holographic dark energy~\cite{Huang2019a, Ebrahimi2020, Astashenok2023} and Barrow holographic dark energy~\cite{Huang2021, Srivastava2021}.

The entropy of the horizon underpins holographic dark energy, and different types of horizon entropy lead to different HDE models~\cite{Wang2017}. Recently, a new HDE model was proposed based on the entropy of an anti-de Sitter black hole, with the Hubble horizon and particle horizon chosen as IR cutoffs~\cite{Nakarachinda2022}. This model has been shown to achieve late-time acceleration and describe the entire evolutionary history of the universe, which is not accomplished in the original HDE model~\cite{Hsu2004}. However, a successful HDE model must also remain stable against perturbations, which is determined by the squared sound speed $v_{s}^{2}$~\cite{Peebles2003}. When the Hubble horizon is chosen as the IR cutoff, the HDE model may become unstable against perturbations~\cite{Tavayef2018, Zadeh2018, Srivastava2021}. Whether the new HDE model with the Hubble horizon as an IR cutoff (NHDEH) can describe the entire evolutionary history of the universe; whether classical instability in this model can be resolved; and if NHDEH can describe the entire evolutionary history of the universe and is stable, whether it can be distinguished from the standard $\Lambda$CDM model. These issues will be explored in this paper.

This paper has four main objectives: to investigate inflation, to explore reheating, to address the unresolved question of classical instability, and to examine the evolution of the universe in NHDEH. The paper is organized as follows: In Section II, we examine inflation in NHDEH. In Section III, we explore reheating in NHDEH. In Section IV, we analyze both the evolution of the universe and the classical instability in NHDEH. Finally, our main conclusions are presented in Section V.

\section{Inflation}

Using the entropy of anti-de Sitter black hole, the energy density of NHDEH is given as~\cite{Nakarachinda2022}
\beq
\rho_{de}=\frac{3}{\kappa^{2}} b^{2} \Big(\frac{1}{L^{2}}+\Lambda\Big)=\frac{3}{\kappa^{2}} b^{2} (H^{2}+\Lambda).\label{rhode}
\eeq
Here, $\kappa^{2}=8\pi G$, $b$ is a dimensionless parameter, and $\Lambda$ represents the cosmological constant, which provides a possibility of obtaining the late-time acceleration of the universe. For $\Lambda=0$, the energy density of NHDEH reduces to that of the original holographic dark energy~\cite{Hsu2004}, which cannot explain the late-time cosmic acceleration or describe the complete evolutionary history of the universe.

We consider a flat Friedmann-Robertson-Walker universe that is homogeneous and isotropic
\beq
ds^{2}=-dt^{2}+a^{2}(t)(dr^{2}+r^{2}d\Omega^{2}),
\eeq
where $a(t)$ is the scale factor with $t$ being cosmic time. The Friedmann equations are given as
\bea
& H^{2}=\displaystyle\frac{\kappa^{2}}{3}\big(\rho_{\phi}+\rho_{de}\big),\label{H2}\\
& 2\dot{H}+3H^{2}=-\kappa^{2}(p_{\phi}+p_{de})\label{H1}.
\eea
Here, $\rho_{\phi}=\frac{1}{2}\dot{\phi}^{2}+V(\phi)$ and $p_{\phi}=\frac{1}{2}\dot{\phi}^{2}-V(\phi)$ represent the energy density and pressure of a scalar field that satisfies
\beq
\ddot{\phi}+3H\dot{\phi}+V_{,\phi}=0.\label{phi2}.
\eeq
And $\rho_{de}$ and $p_{de}$ are the energy density and pressure of NHDEH, which satisfy
\beq
\dot{\rho}_{de}+3H(\rho_{de}+p_{de})=0,\label{cde}
\eeq
with $p_{de}=\omega_{de}\rho_{de}$ where $\omega_{de}$ is the equation of state parameter.

\subsection{Holographic inflation}

To analyze holographic inflation in NHDEH, we consider inflation driven by holographic dark energy while ignoring the effect of the scalar field $\phi$. Thus, the Friedmann equations ~(\ref{H2}) and ~(\ref{H1}) become
\bea
& H^{2}=\displaystyle\frac{\kappa^{2}}{3} \rho_{de},\\
& 2\dot{H}+3H^{2}=-\kappa^{2} p_{de},
\eea
which give the relation
\beq
\dot{H}=-\frac{3}{2} \big[(1+b^{2} \omega_{de})H^{2}+b^{2} \Lambda \omega_{de} \big].\label{H11}
\eeq
Substituting Eq.~(\ref{rhode}) into Eq.~(\ref{cde}), we obtain
\beq
\omega_{de}=-\frac{2\dot{H}+3H^{2}+3\Lambda}{3(H^{2}+\Lambda)}.\label{ode}
\eeq
Using the relation given in Eq.~(\ref{ode}), Eq.~(\ref{H11}) is expressed as
\beq
\dot{H}=\frac{3}{2}\Big(-H^{2}+\frac{b^{2}\Lambda}{1-b^{2}}\Big).\label{H111}
\eeq

\begin{figure*}[htp]
\begin{center}
\includegraphics[width=0.6\textwidth]{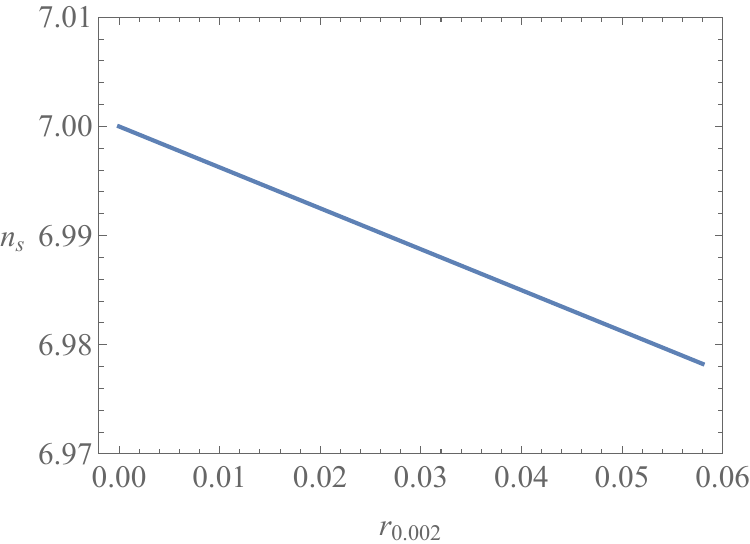}
\caption{\label{Fig1} Relation between $n_{s}$ and $r_{0.002}$.}
\end{center}
\end{figure*}

To study the inflation model, the slow-roll parameters $\epsilon_n$, defined by the Hubble parameters $H$ and $\dot{H}$, can be directly calculated~\cite{Martin2014}. Then, by utilizing Eq.~(\ref{H111}), we derive the first and second slow-roll parameters and express them as
\bea
& \epsilon_{1}=-\displaystyle\frac{\dot{H}}{H^{2}}=\frac{3}{2}\big(1-\frac{b^{2}}{1-b^{2}}\frac{\Lambda}{H^{2}}\big),\\
& \epsilon_{2}=\displaystyle\frac{\dot{\epsilon}_{1}}{H \epsilon_{1}}=-3\frac{b^{2}}{1-b^{2}}\frac{\Lambda}{H^{2}}.
\eea
Now, utilizing these slow-roll parameters, we can calculate the values of the inflationary observables, namely the scalar spectral index of the curvature perturbations $n_{s}$ and the tensor-to-scalar ratio $r$~\cite{Martin2014}, which can be expressed as
\bea
& n_{s}=1-2\epsilon_{1}-2\epsilon_{2}=-2+9\displaystyle\frac{b^{2}}{1-b^{2}}\frac{\Lambda}{H^{2}},\label{nsh}\\
& r=16\epsilon_{1}=24\big(1-\displaystyle\frac{b^{2}}{1-b^{2}}\frac{\Lambda}{H^{2}}\big).\label{rh}
\eea
Substituting Eq.~(\ref{rh}) into Eq.~(\ref{nsh}), we obtain the relation
\beq
n_{s}=7-\frac{3}{8}r,
\eeq
which is plotted in Fig.~(\ref{Fig1}). From this figure, we can observe that $n_{s}$ is larger than $6.98$, which exceeds the Planck 2018 results of $n_{s}=0.9668 \pm 0.0037$ for $r_{0.002}<0.058$ ~\cite{Planck2020}. Therefore, in NHDEH, observations do not support holographic inflation. It is worth noting that holographic inflation is realized in other holographic dark energy model with a different IR cutoff ~\cite{Nojiri2019, Oliveros2019,Mohammadi2021}.

\subsection{Slow-roll inflation}

In the previous subsection, we analyzed holographic inflation in NHDEH and found that it is not supported by the Planck 2018 results. We will discuss whether the slow-roll inflation in NHDEH can be supported by observations in this subsection. By applying the slow-roll conditions $\frac{1}{2}\dot{\phi}^{2} \ll V(\phi)$ and $\mid \ddot{\phi} \mid \ll \mid 3 H \dot{\phi} \mid$, Eqs.~(\ref{H2}) and ~(\ref{phi2}) can be expressed as
\bea
& H^{2} \approx \displaystyle\frac{\kappa^{2}}{3}\frac{1}{1-b^{2}}V+\frac{b^{2}}{1-b^{2}}\Lambda,\label{H20}\\
& 3H\dot{\phi}+V_{,\phi} \approx 0.\label{phi20}
\eea
Taking the derivative of Eq.~(\ref{H20}) yields
\beq
\dot{H}=-\frac{\kappa^{2} V_{,\phi}^{2}}{18(1-b^{2}) H^{2}}.\label{H21}
\eeq

\begin{figure*}[htp]
\begin{center}
\includegraphics[width=0.45\textwidth]{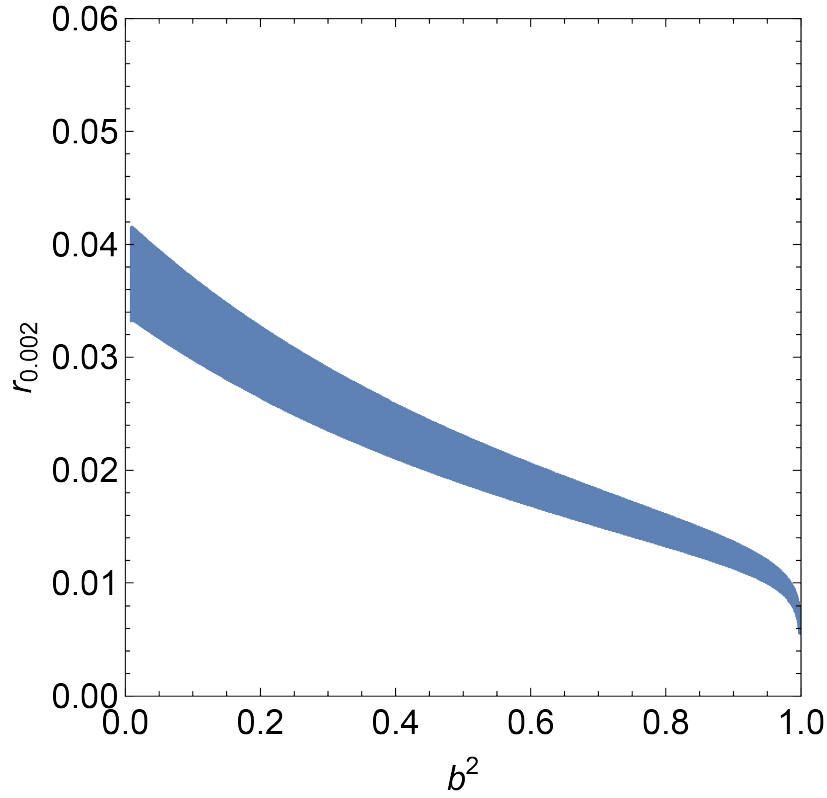}
\includegraphics[width=0.45\textwidth]{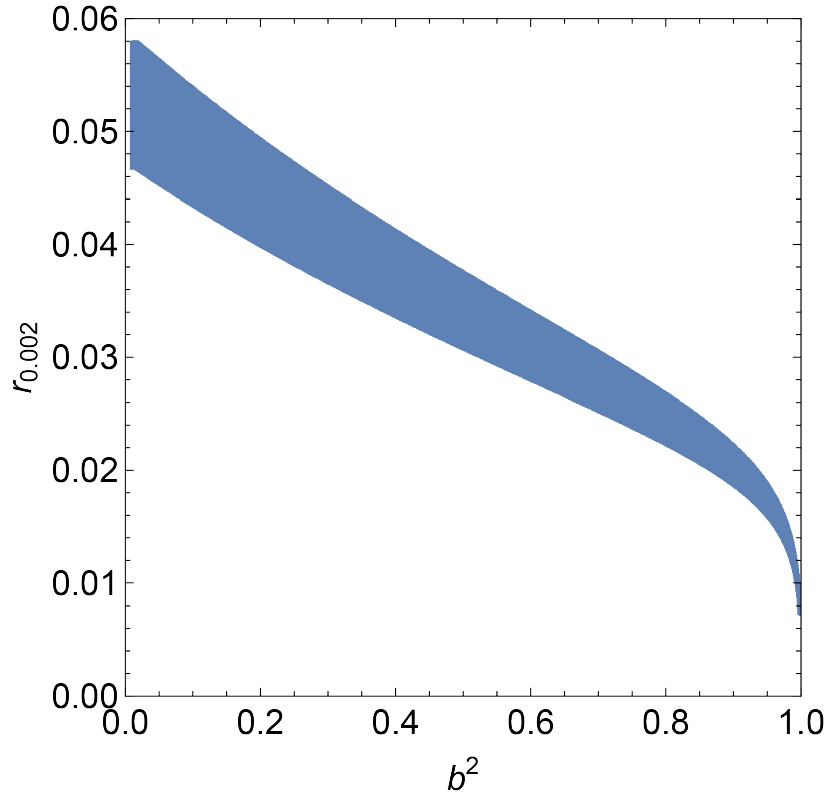}
\includegraphics[width=0.45\textwidth]{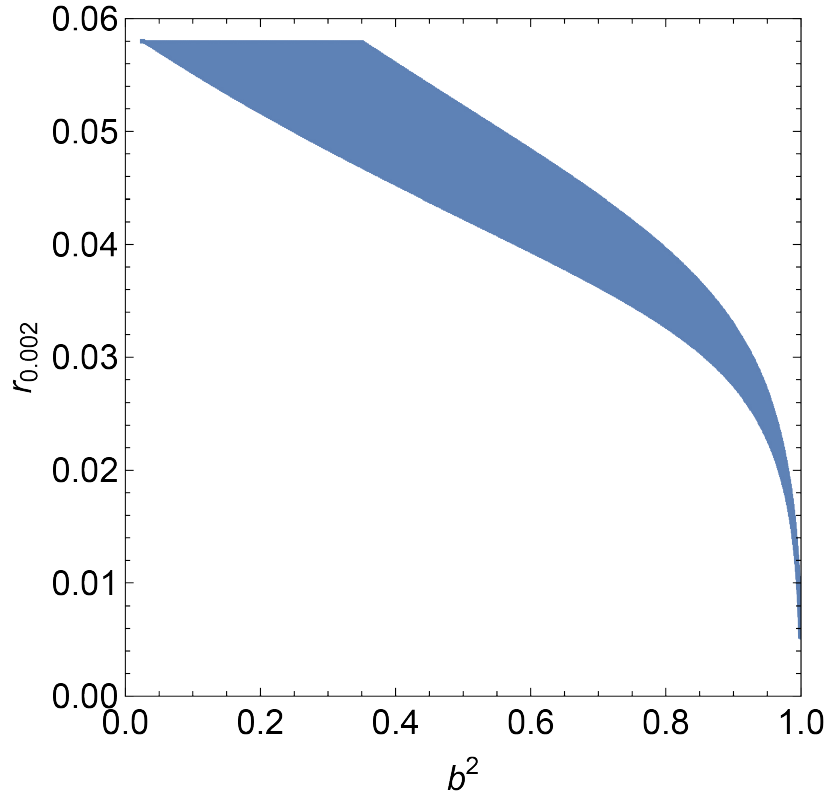}
\includegraphics[width=0.45\textwidth]{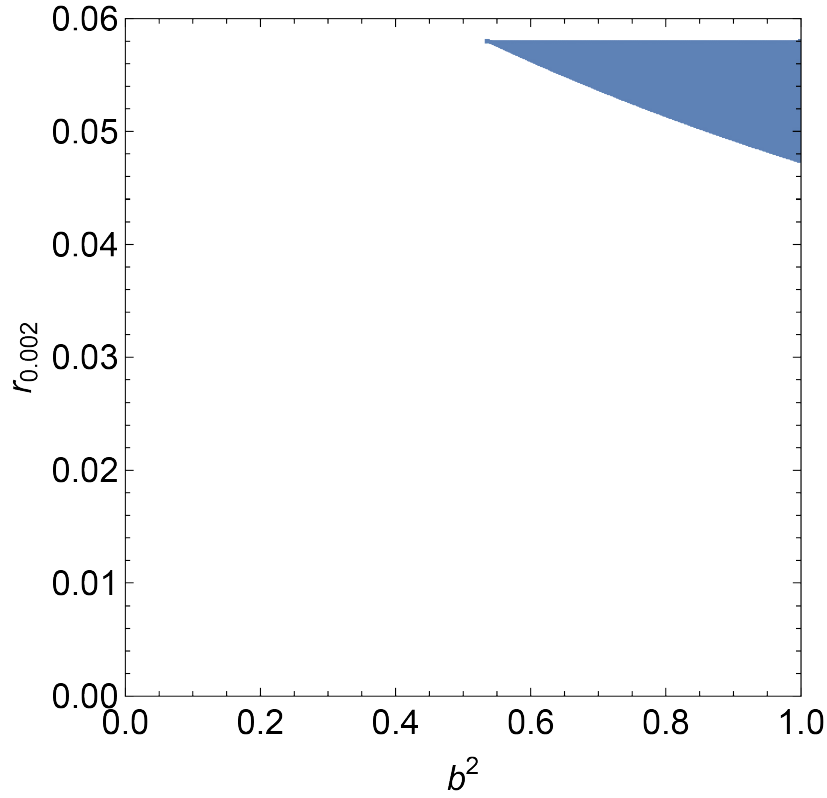}
\caption{\label{Fig2} Relation between $r_{0.002}$ and $b^{2}$ for $n=\frac{1}{3}$, $\frac{1}{2}$, $\frac{2}{3}$, and $1$, respectively.}
\end{center}
\end{figure*}

To study slow-roll inflation in NHDEH, we can consider inflation driven by the scalar field and assume that the existence of holographic dark energy does not affect the inflation~\cite{Chen2007, Wang2017}. Thus, the scalar spectral index of the curvature perturbations $n_{s}$ and the tensor-to-scalar ratio $r$ take the form~\cite{Chen2007}
\bea
& n_{s} = 1-8\epsilon+2\delta+2\sigma,\\
& r = 16\epsilon, \label{rrr}
\eea
with
\bea
& \epsilon = -\displaystyle\frac{\dot{H}}{H^{2}},\\
& \delta = -\displaystyle\frac{\ddot{\phi}}{H \dot{\phi}},\\
& \sigma = \displaystyle\frac{\kappa^{2} \dot{\phi}^{2}}{H^{2}}.
\eea
By utilizing Eqs.~(\ref{H20}), ~(\ref{phi20}) and ~(\ref{H21}), we can rewrite $\epsilon$, $\delta$ and $\sigma$ as
\bea
& \epsilon =\displaystyle\frac{\kappa^{2}}{2(1-b^{2})}\Big(\frac{V_{,\phi}}{3H^{2}}\Big)^{2},\label{epsilon1}\\
& \delta = -\epsilon+\displaystyle\frac{V_{,\phi\phi}}{3H^{2}},\\
& \sigma = \kappa^{2} \Big(\displaystyle\frac{V_{,\phi}}{3H^{2}}\Big)^{2}.
\eea
To calculate the values of $n_{s}$ and $r$, we consider that the potential $V(\phi)$ takes the form
\beq
V=V_{0}\phi^{n},\label{Vp}
\eeq
which is known as the chaotic potential. Here, both $V_{0}$ and $n$ are positive parameters. For convenience, we set $V_{0} = \Lambda$ and obtain
\bea
& n_{s} = 1+\displaystyle\frac{(2b^{4}+b^{2}-3) \kappa^{2} n^{2} \phi^{2n-2}}{(3b^{2}+\kappa^{2} \phi^{n})^{2}}+\frac{2(1-b^{2})(n-1)n \phi^{n-2}}{3b^{2}+\kappa^{2} \phi^{n}},\label{ns1}\\
& r = \displaystyle\frac{8(1-b^{2})\kappa^{2} n^{2} \phi^{2n-2}}{(3b^{2}+\kappa^{2} \phi^{n})^{2}}.\label{r1}
\eea
To obtain the relationship between $n_{s}$ and $r$, we need to solve for $\phi$ from Eq.~(\ref{r1}) and then substitute $\phi$ into Eq.~(\ref{ns1}). However, since the expression of $\phi$ depends on the values of $n$ and $b^2$ and is very complicated, we can only illustrate their relationship through figures. In Fig.~(\ref{Fig2}), considering the restricted condition $n_{s}=0.9668 \pm 0.0037$ for $r_{0.002}<0.058$~\cite{Planck2020}, we have plotted the relation between $r_{0.002}$ and $b^{2}$ for $n=\frac{1}{3}$, $\frac{1}{2}$, $\frac{2}{3}$, and $1$, respectively. From these figures, we can see that the slow-roll inflation can be realized by choosing the appropriate values for $n$ and $b^{2}$, and as $n$ decreases, the range of values for $b^{2}$ increases.

Although slow-roll inflation can be realized in NHDEH, it is uncertain whether we can achieve the required e-folds number $N$, i.e., $50 < N < 70$. To answer this question, we will conduct a further analysis of slow-roll inflation in NHDEH. According to the definition of the e-folds number $N$, we can write it as
\beq
N = \ln \Big( \frac{a_e}{a_i} \Big) = \int^{t_e}_{t_i} H dt = -\int^{\phi_{e}}_{\phi_i} \frac{3H^{2}}{V_{,\phi}}d \phi.\label{N01}
\eeq
Here, Eq.~(\ref{phi20}) is used, the subscript $_{e}$ denotes the time when inflation ended, while the subscript $_{i}$ represents the time when inflation began. Then, substituting Eqs.~(\ref{phi20}) and ~(\ref{Vp}) into Eq.~(\ref{N01}), we get
\beq
N = \frac{1}{2(1-b^{2})n}\Big[\kappa^{2}(\phi^{2}_{e}-\phi^{2}_{i})-\frac{6b^{2}}{n-2}(\phi^{2-n}_{e}-\phi^{2-n}_{i})\Big].\label{N1}
\eeq
Since the first slow-roll parameter at the end of inflation is known to take the value $\epsilon_{e} \simeq 1$, we can solve the expression for $\phi_{e}$ using this relation. By substituting $\phi_{e}$ into Eq.~(\ref{N1}), the expression for $\phi_{i}$ can be obtained. Then, combining Eqs.~(\ref{ns1}), ~(\ref{r1}), and $\phi_{i}$, we can obtain the expressions for $n_{s}$ and $r$. Similar to the situation in holographic inflation, the expressions for $\phi_{e}$, $\phi_{i}$, $n_{s}$, and $r$ depend on the values of $n$, $b^{2}$, and $N$ and cannot be expressed analytically. We will use figures to show their relationships.

\begin{figure*}[htp]
\begin{center}
\includegraphics[width=0.45\textwidth]{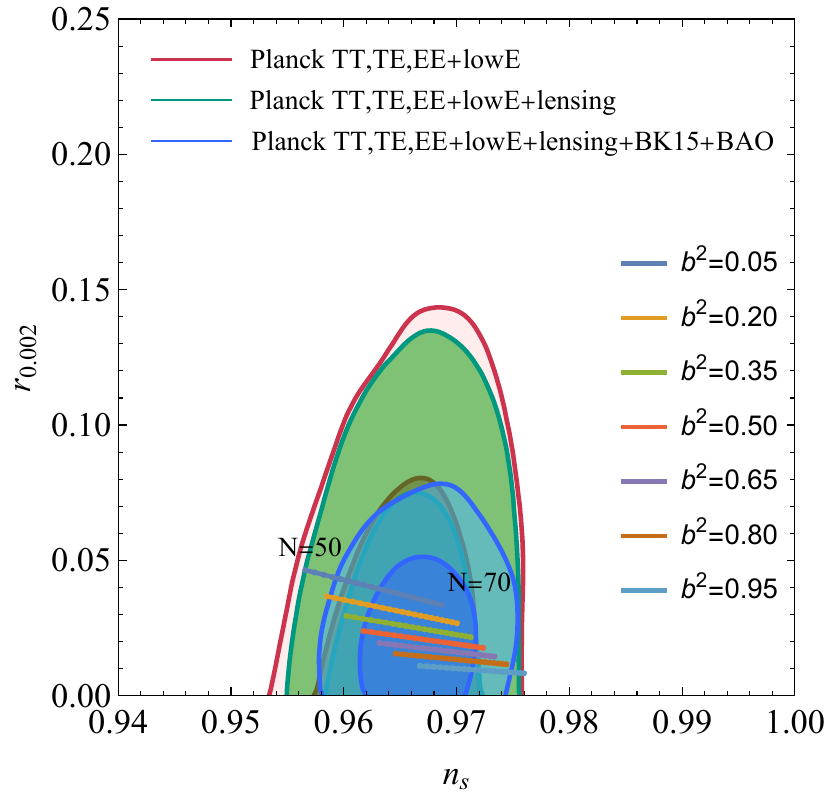}
\includegraphics[width=0.45\textwidth]{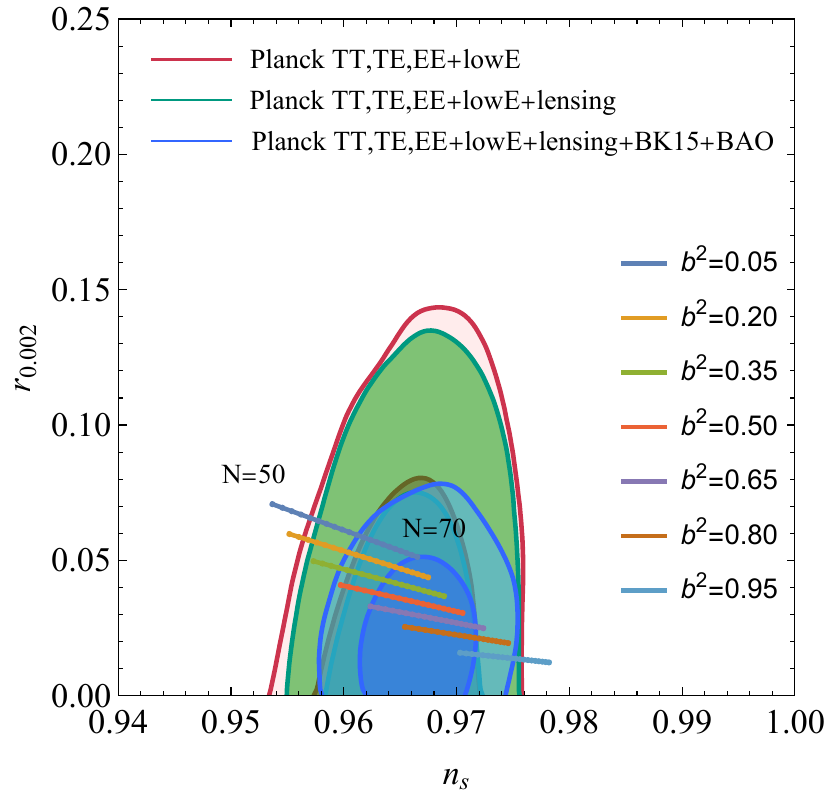}
\includegraphics[width=0.45\textwidth]{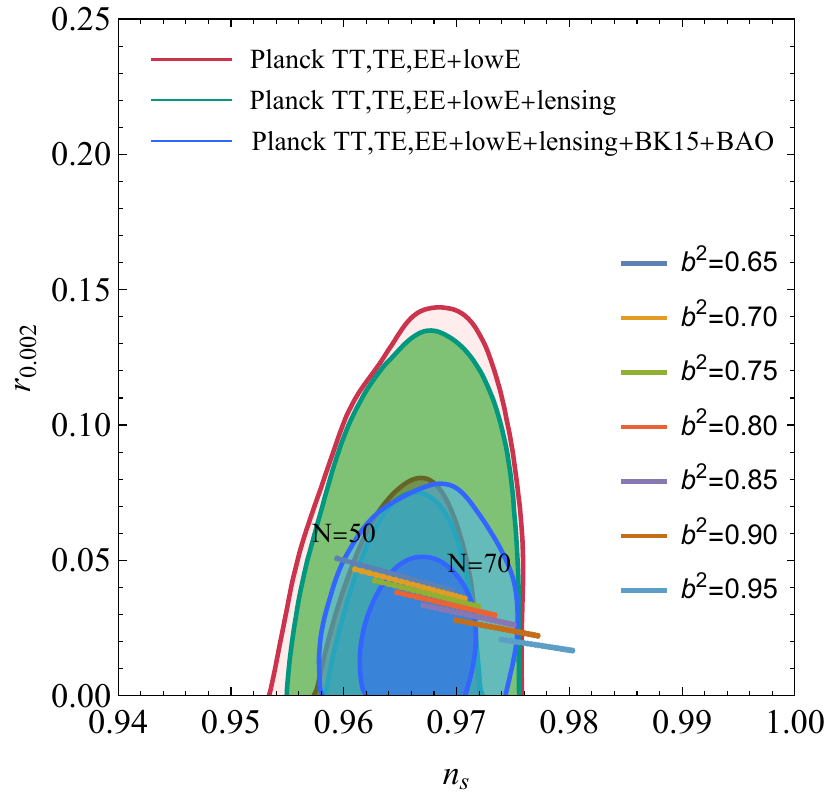}
\includegraphics[width=0.45\textwidth]{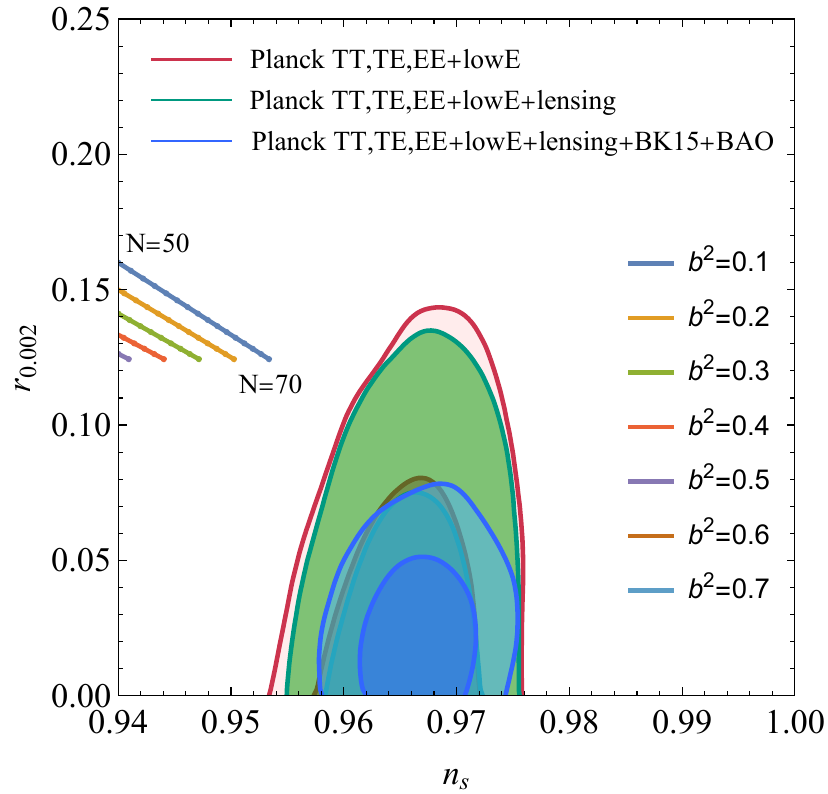}
\caption{\label{Fig3} Predictions of the chaotic potential ~(\ref{Vp}) in $r_{0.002}-n_{s}$ plane. These figures are depicted for $n=\frac{1}{3}$, $\frac{1}{2}$, $\frac{2}{3}$, and $1$, respectively.}
\end{center}
\end{figure*}

Fixing the values of $n$ and $b^{2}$, we plot the predictions of the chaotic potential ~(\ref{Vp}) in the $r_{0.002}-n_{s}$ plane in Fig.~(\ref{Fig3}), where we overlay our analytical results with Planck 2018 data. These figures show that slow-roll inflation can be realized for $n=\frac{1}{3}$, $\frac{1}{2}$, and $\frac{2}{3}$ by selecting the appropriate values for $b^{2}$ and the e-folds number $N$. The value of $r_{0.002}$ decreases as $N$ increases from $50$ to $70$ and as $b^{2}$ increases, while $n_s$ decreases as $N$ decreases from $70$ to $50$ and as $b^{2}$ decreases. With this comparison to the Planck 2018 data, the results favor the cases where $n=\frac{1}{3}$ with $b^{2}=0.50$, and $n=\frac{1}{2}$ with $b^{2}=0.65$. It is clear that an increased value of $n$ may result in a mismatch with the observations. In addition, using the binned Pantheon, cosmic chronometers, and baryon acoustic oscillations datasets, the observed value for $b^{2}$ is $b^{2}=0.01$~\cite{Nakarachinda2022}. This value is not excluded when $n=\frac{1}{3}$. Notably, the chaotic potential~(\ref{Vp}) is ruled out by the Planck 2018 data in the context of standard slow-roll inflation~\cite{Planck2020a}. However, slow-roll inflation is realized within NHDEH and is consistent with the Planck 2018 data.

\section{Reheating}

In the previous section, we have analyzed holographic inflation and slow-roll inflation in NHDEH. After inflation ends, i.e., $\epsilon_{e} \simeq 1$, the universe reheats, after which the evolution of the hot big bang universe begins. During the reheating phase, the inflaton begins to oscillate around the bottom of the potential and is expected to decay into other particles~\cite{Bassett2006, Cai2015}. To describe the reheating phase, the reheating temperature $T_{reh}$, the effective equation of state parameter $\omega_{reh}$ and the number of reheating e-folds $N_{reh}$ are introduced~\cite{Dai2014, Cook2015, Munoz2015, Goswami2018, Zhou2022, Yadav2024, Zhang2021, Zhang2023}. In this section, we will discuss whether $T_{reh}$ and $N_{reh}$ produced during reheating in NHDEH can be supported by observations.

Combining Eqs.~(\ref{H20}), ~(\ref{phi20}), and ~(\ref{epsilon1}), the energy density and Hubble parameter during inflation can be written as
\bea
\rho_{\phi} = \frac{3V}{3-\epsilon} + \frac{3 b^{2} \Lambda \epsilon}{\kappa^{2}(3-\epsilon)},\\
H^{2} = \frac{\kappa^{2} V + 3 b^{2} \Lambda}{(1-b^{2})(3-\epsilon)}.\label{H2r}
\eea
Here, the potential $V$ is given in Eq.~(\ref{Vp}). After inflation ends, the slow-roll parameter $\epsilon$ satisfies $\epsilon_{e} \simeq 1$. Hence, the energy density at the end of inflation becomes
\beq
\rho_{e} = \frac{3}{2}\Big( V_{e}+\frac{b^{2} V}{\kappa^{2}} \Big).\label{rhoe}
\eeq

Similar to the definition of the e-folds number during inflation, the number of e-folds between the time when a mode $k$ crosses the Hubble horizon and the end of inflation is defined as
\beq
\Delta N_{k} = \ln \Big( \frac{a_{e}}{a_{k}} \Big) = \int^{\phi_{e}}_{\phi_{k}} \frac{H}{\dot{\phi}} d \phi,
\eeq
where the subscript $_{k}$ denotes the time when the mode $k$ crosses the Hubble radius. And the number of reheating e-folds $N_{reh}$ is given by
\beq
\Delta N_{reh} = \ln \Big( \frac{a_{reh}}{a_{e}} \Big),
\eeq
where the subscript $_{reh}$ denotes the end of reheating. Since $N_{reh}$ is defined as the number of e-folds from the end of inflation to the beginning of the radiation-dominated era, it encodes both the epoch of reheating and the subsequent thermalization process. The subsequent evolution of the universe is governed by its energy density
\beq
\rho_{reh} = \frac{\pi^{2}}{30} g_{reh} T^{4}_{reh},\label{rhoreh0}
\eeq
where $g_{reh}$ represents the effective number of relativistic species and $T_{reh}$ is the temperature at the end of reheating. Then, considering the continuity equation, we have
\beq
\dot{\rho} + 3H(\rho + p) = \dot{\rho} + 3H (1 + \omega_{reh}) \rho = 0.
\eeq
In view of this equation, we get~\cite{Goswami2018}
\beq
\rho_{reh} = \rho_{e} e^{-3 N_{reh} (1 + \overline{\omega}_{reh})}.\label{rhoreh1}
\eeq
Here, $\overline{\omega}_{reh}$ is the average equation of state parameter during reheating~\cite{Martin2015}. Eq.~(\ref{rhoreh1}) can be rewritten as
\beq
\frac{a_{reh}}{a_{e}} = e^{N_{reh}} = \Big( \frac{\rho_{reh}}{\rho_{e}} \Big)^{-\frac{1}{3(1 + \overline{\omega}_{reh})}}.\label{aareh1}
\eeq

In order to relate the observed wavenumber of any physical scale today $\frac{k}{a_{0}}$ to its value at the time of Hubble crossing during inflation $\frac{k}{a_{k}}$, we have~\cite{Goswami2018}
\beq
\frac{k}{a_{k}} = \frac{k}{a_{0}}\frac{a_{0}}{a_{k}} = \frac{k}{a_{0}}\frac{a_{0}}{a_{eq}}\frac{a_{eq}}{a_{reh}}\frac{a_{reh}}{a_{e}}\frac{a_{e}}{a_{k}} = \frac{k}{a_{0}} (1+z_{eq}) \Big( \frac{\rho_{reh}}{\rho_{eq}} \Big)^{\frac{1}{4}} e^{N_{reh}} e^{\Delta N_{k}},\label{kak1}
\eeq
where the subscript $_{eq}$ denotes the epoch of matter-radiation equality. Combining Eqs.~(\ref{rhoe}), ~(\ref{aareh1}), and ~(\ref{kak1}), we obtain
\beq
\frac{k}{a_{k}} = H_{k} = \frac{k}{a_{0}} (1+z_{eq}) \rho_{reh}^{\frac{3\overline{\omega}_{reh}-1}{12(1+\overline{\omega}_{reh})}} \rho_{eq}^{-\frac{1}{4}} \Big[ \frac{3}{2} \Big( V_{e} + \frac{b^{2} \Lambda}{\kappa^{2}} \Big) \Big]^{\frac{1}{3(1+\overline{\omega}_{reh})}} e^{\Delta N_{k}}.\label{kak2}
\eeq
Substituting Eq.~(\ref{rhoreh0}) into ~(\ref{kak2}), the expression for $T_{reh}$ can be written as
\bea
\ln(T_{reh}) &&= \frac{3(1+\overline{\omega}_{reh})}{3\overline{\omega}_{reh} -1} \Big[ \ln(H_{k})-\ln\Big( \frac{k}{a_{0}} \Big)-\ln(1+z_{eq})-\Delta N_{k}+\ln\Big( \rho_{eq}^{\frac{1}{4}} \Big) \Big] \nonumber\\
&&-\frac{1}{3\overline{\omega}_{reh}-1}\ln\Big[ \frac{3}{2}\Big( V_{e}+\frac{b^{2} \Lambda}{\kappa^{2}} \Big) \Big]-\frac{1}{4}\ln\Big( \frac{\pi^{2}}{30} g_{reh} \Big).\label{Treh}
\eea
Combining Eqs.~(\ref{rhoreh0}), ~(\ref{aareh1}), and ~(\ref{Treh}), we can obtain the expression for $N_{reh}$ as follows
\bea
N_{reh} &&= \frac{4}{3\overline{\omega}_{reh}-1}\Big[ \ln\Big( \frac{k}{a_{0}} \Big)+\ln(1+z_{eq})+\Delta N_{k}-\ln(H_{k})-\ln\Big( \rho_{eq}^{\frac{1}{4}} \Big) \Big] \nonumber\\
&&+\frac{1}{3\overline{\omega}_{reh}-1}\ln\Big[ \frac{3}{2}\Big( V_{e}+\frac{b^{2} \Lambda}{\kappa^{2}} \Big) \Big].\label{Nreh}
\eea
Eqs.~(\ref{Treh}) and ~(\ref{Nreh}) are the two key relationships that can be used to parameterize reheating in slow-roll inflationary models. To study the constraints on reheating from the CMB data, we consider a pivot scale $k_{*}$ which corresponds to the time when the observable CMB scales exit the Hubble radius $H_{*}$ during inflation. For the pivot scale $k_{*}$, we take the value $\frac{k_{*}}{a_{0}}=0.05Mpc^{-1}$ used by the Planck collaboration. For $k=k_{*}$, the amplitude of scalar power spectra is given in terms of $H_{*}$ as
\beq
A_{s} = P_{\zeta}(k_{*}) = \frac{\kappa^{2} H^{2}_{*}}{8 \pi^{2} \epsilon_{*}},
\eeq
which gives the relation
\beq
H_{*} = \frac{\pi}{\kappa}\sqrt{8 A_{s} \epsilon_{*}}.\label{H1s}
\eeq
The number of e-folds remaining after the pivot scale $k_{*}$ crosses the Hubble radius is given by
\bea
\Delta N_{*} = \int^{\phi_{*}}_{\phi_{e}} \frac{H}{\dot{\phi}} d \phi = \frac{\kappa^{2}}{2(1-b^{2})n}(\phi^{2}_{*}-\phi^{2}_{e})+\frac{3 b^{2}}{(1-b^{2})(2-n)n}(\phi^{2-n}_{*}-\phi^{2-n}_{e}),\label{N1s}
\eea
where $\phi_{e}$ can be solved numerically using Eqs.~(\ref{H20}), ~(\ref{phi20}), and ~(\ref{epsilon1}) under the condition $\epsilon_{e} \simeq 1$. Then, using Eqs.~(\ref{H20}), ~(\ref{phi20}), and ~(\ref{epsilon1}), the spectral index $n_{s}$~(\ref{ns1}) can be rewritten as
\beq
n_{s} = 1-\Big[ 2\Big( 1+2 b^{2}+\frac{2}{n} \Big)-12\frac{b^{2}(n-1)}{\kappa^{2} n} \Big]\epsilon_{*}.\label{ns1s}
\eeq
Using Eq.~(\ref{H1s}), we obtain the relation
\beq
V_{e}+\frac{b^{2} \Lambda}{\kappa^{2}} = \frac{3 H^{2}_{*} \phi^{n}_{e}}{\kappa^{2} \phi^{n}_{*}}+\frac{b^{2}}{\kappa^{2}}\Big[ \Lambda-3(\Lambda+H^{2}_{*})\frac{\phi^{n}_{e}}{\phi^{n}_{*}} \Big].\label{Ve1b2}
\eeq

\begin{figure*}[htp]
\begin{center}
\includegraphics[width=0.45\textwidth]{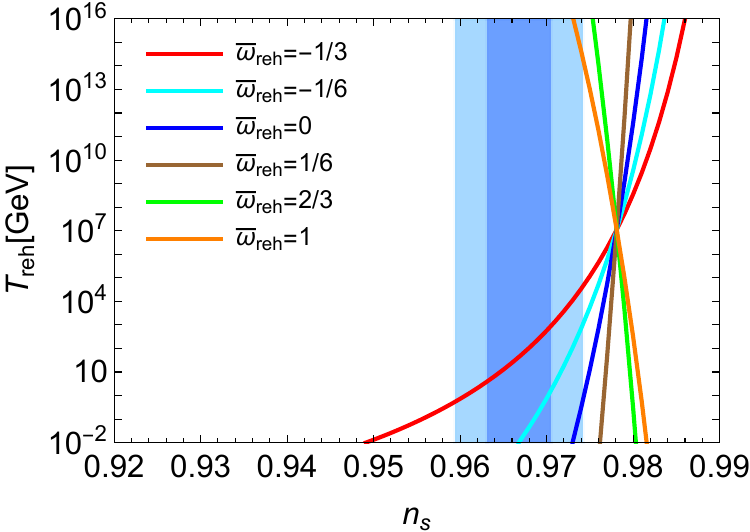}
\includegraphics[width=0.45\textwidth]{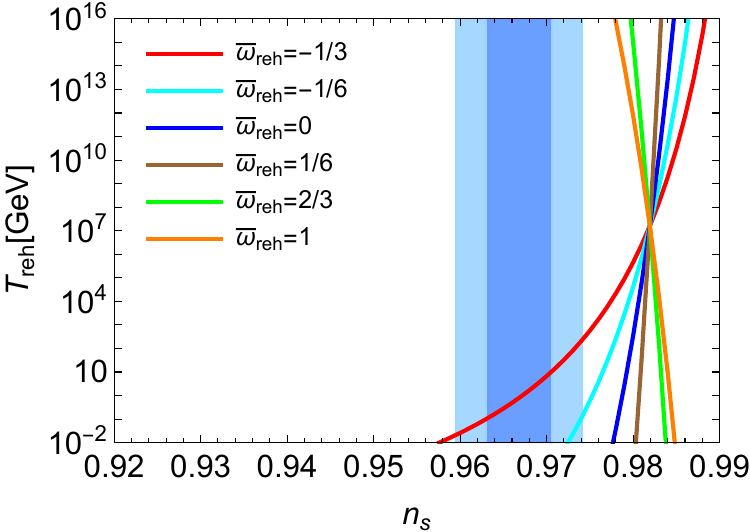}
\caption{\label{Fig3a} Reheating temperature $T_{reh}$ versus $n_{s}$ for NHDEH. The left panel is plotted for $n=\frac{1}{3}$, while the right one is for $n=\frac{1}{2}$.}
\end{center}
\end{figure*}

\begin{figure*}[htp]
\begin{center}
\includegraphics[width=0.45\textwidth]{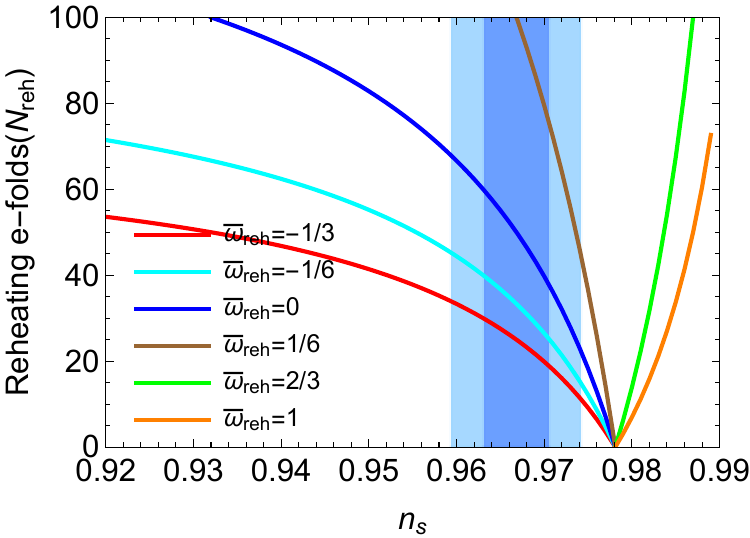}
\includegraphics[width=0.45\textwidth]{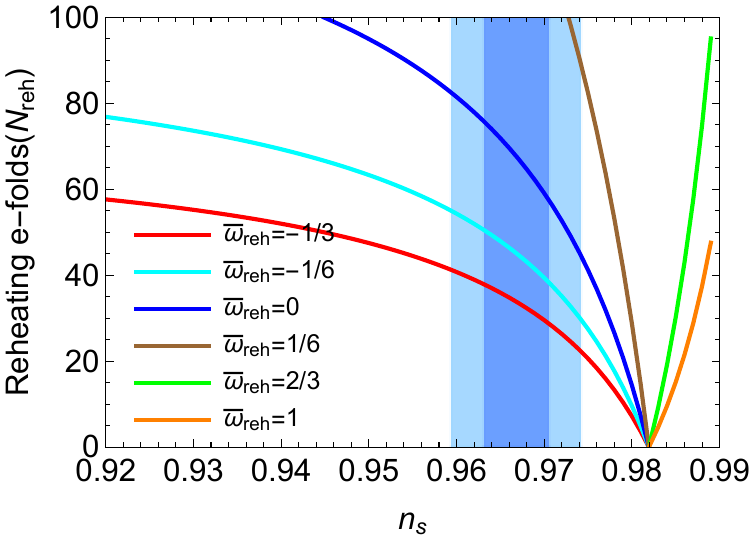}
\caption{\label{Fig3b} Number of reheating e-folds $N_{reh}$ versus $n_{s}$ for NHDEH. The left panel is plotted for $n=\frac{1}{3}$, while the right one is for $n=\frac{1}{2}$.}
\end{center}
\end{figure*}

Now, using Eqs.~(\ref{H1s}), ~(\ref{N1s}), ~(\ref{ns1s}), and ~(\ref{Ve1b2}), the reheating temperature $T_{reh}$~(\ref{Treh}) and the number of reheating e-folds $N_{reh}$~(\ref{Nreh}) can be expressed as functions of $\overline{\omega}_{reh}$ and $n_{s}$ for the pivot scale. Then, taking $\hbar = c = 1$, $g_{reh}=100$, $\kappa=4.106 \times 10^{-9} GeV^{-1}$  and using Planck 2018 results of $A_{s}=2.105 \times 10^{9}$, $z_{eq}=3387$, and $H_{0}=67.66 km s^{-1}Mpc^{-1}$~\cite{Planck2020}, we have plotted the figures for $T_{reh}$ and $N_{reh}$ versus $n_{s}$, which are shown in Figs.~(\ref{Fig3a}) and ~(\ref{Fig3b}), respectively. Here, we adopt $b^{2}=0.01$~\cite{Nakarachinda2022}. The blue and light blue shaded regions correspond to the $1\sigma$ and $2\sigma$ bounds on $n_{s}$ from Planck 2018 data (TT,TE,EE+lowE+lensing+BK15+BAO)~\cite{Planck2020}.

From Figs.~(\ref{Fig3a}) and ~(\ref{Fig3b}), it can be observed that the instantaneous reheating is excluded by Planck results in NHDEH, as the corresponding spectral index $n_{s}$ exceeds the $2\sigma$ bound on $n_{s}$ from Planck data. In Fig.~(\ref{Fig3a}), within the $2\sigma$ bound on $n_{s}$ from Planck results, the data favor a negative $\overline{\omega}_{reh}$, and the possible allowed values of reheating temperature $T_{reh}$ are greater than $10^{-2}GeV$. Curves for $\overline{\omega}_{reh}=-\frac{1}{3}$ predict the allowed values of reheating temperature $T_{reh}$ to be $4.43 \times 10^{-1} GeV$ to $1.17 \times 10^{5} GeV$ for $n=\frac{1}{3}$ and $1.72 \times 10^{-2} GeV$ to $5.68 \times 10^{2} GeV$ for $n=\frac{1}{2}$. Fig.~(\ref{Fig3b}) shows that the number of reheating e-folds $N_{reh}$ are greater than $9$ for $n=\frac{1}{3}$ and greater than $20$ for $n=\frac{1}{2}$ within the $2\sigma$ bound on $n_{s}$ from Planck results. Curves for $\overline{\omega}_{reh}=-\frac{1}{3}$ predict $9 < N_{reh} < 35$ for $n=\frac{1}{3}$ and $20 < N_{reh} < 42$ for $n=\frac{1}{2}$. Thus, under the constraint $b^{2}=0.01$~\cite{Nakarachinda2022}, the results from reheating favor the cases $n=\frac{1}{3}$ and $n=\frac{1}{2}$.

For the case $b=0$ and $n=2$, NHDEH reduces to the quadratic chaotic inflationary model $V=\frac{1}{2}m^{2}\phi^{2}$, and our results for $T_{reh}$ and $N_{reh}$ will reduce to those in ~\cite{Dai2014, Goswami2018, Yadav2024}.

\section{Evolution of universe}

In the previous section, we discussed reheating in NHDEH. After reheating ends, the universe evolves from a radiation-dominated epoch, transitions to a matter-dominated epoch, and eventually enters an epoch dominated by dark energy. To discuss the evolution of the universe after inflation, we assume that the scalar field $\phi$ decays into radiation and pressureless matter. Thus, the first Friedmann equation ~(\ref{H2}) becomes
\beq
H^{2}=\displaystyle\frac{\kappa^{2}}{3}\big(\rho_{r}+\rho_{m}+\rho_{de}\big),\label{H22}\\
\eeq
where $\rho_{r}$ and $\rho_{m}$ denote the energy density for radiation and pressureless matter, respectively, and satisfy
\bea
& \dot{\rho}_{r} + 4H \rho_{r}=0,\label{rhor}\\
& \dot{\rho}_{m} + 3H \rho_{m}=0.\label{rhom}
\eea
Introducing the dimensionless variables
\beq
\Omega_{r}=\frac{\kappa^{2} \rho_{r}}{3H^{2}}, \qquad \Omega_{m}=\frac{\kappa^{2} \rho_{m}}{3H^{2}}, \qquad \Omega_{de}=\frac{\kappa^{2} \rho_{de}}{3H^{2}},
\eeq
the Friedmann equation~(\ref{H22}) can be written in the form
\beq
\Omega_{r}+\Omega_{m}+\Omega_{de}=1.\label{O1}
\eeq
Combining Eqs.~(\ref{cde}), ~(\ref{H22}), ~(\ref{rhor}), ~(\ref{rhom}), and ~(\ref{O1}), we get
\beq
\frac{\dot{H}}{H^{2}}=\frac{1}{2}\Big[\Omega_{m}+(1-3\omega_{de})\Omega_{de}\Big]-2.\label{HH2}
\eeq
And the deceleration parameter $q$ can be calculated as
\beq
q=-1-\frac{\dot{H}}{H^{2}}.
\eeq
Using Eqs.~(\ref{cde}), ~(\ref{rhom}), ~(\ref{O1}), and ~(\ref{HH2}), the dynamical equations for the density parameters $\Omega_{m}$ and $\Omega_{de}$ can be written according to the autonomous system as
\bea
\Omega'_{m}=[(3\omega_{de}-1)\Omega_{de}-\Omega_{m}+1]\Omega_{m},\label{Omm}\\
\Omega'_{de}=[(3\omega_{de}-1)(\Omega_{de}-1)-\Omega_{m}]\Omega_{de}.\label{Omde}
\eea
Here, $\Omega'$ represents $\frac{d\Omega}{d(lna)}$, and the equation of state parameter $\omega_{de}$ can be solved by substituting Eq.~(\ref{rhode}) into Eq.~(\ref{cde}) and expressed as
\beq
\omega_{de}=-\frac{(\Omega_{m}+\Omega_{de}-4)b^{2}+3\Omega_{de}}{3(1-b^{2})\Omega_{de}}.
\eeq
To discuss the stability of NHDEH, we need to analyze its squared sound speed
\beq
v^{2}_{s}=\frac{dp_{de}}{d\rho_{de}}=\frac{\dot{p}_{de}}{\dot{\rho}_{de}}=\frac{\rho_{de}}{\dot{\rho}_{de}}\dot{\omega}_{de}+\omega_{de}.
\eeq
For $v^{2}_{s}>0$, this model can be stable against perturbations. Otherwise, it is unstable.

\subsection{Evolution and stability}

\begin{figure*}[htp]
\begin{center}
\includegraphics[width=0.45\textwidth]{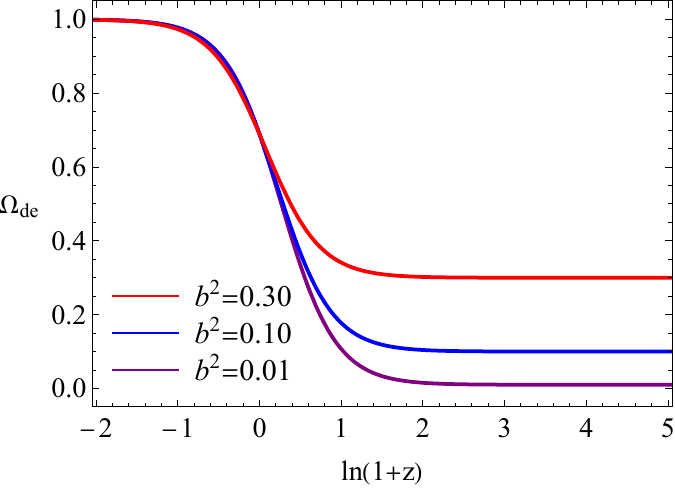}
\includegraphics[width=0.45\textwidth]{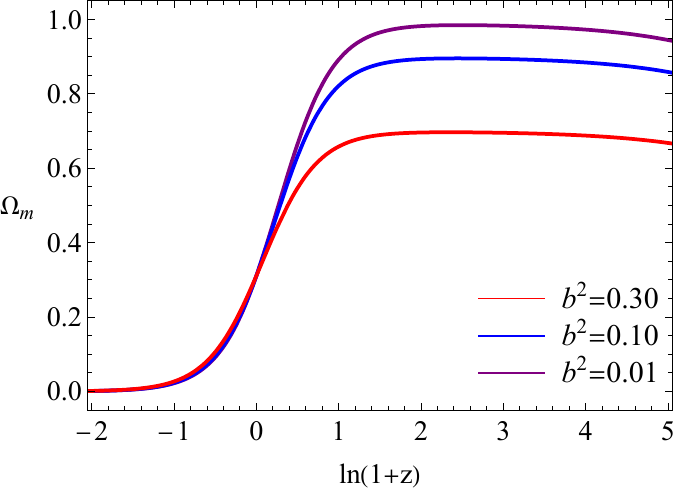}
\caption{\label{Fig4} Evolution curves of $\Omega_{de}$ and $\Omega_{m}$ versus redshift parameter $ln(1+z)$ for NHDEH.}
\end{center}
\end{figure*}

\begin{figure*}[htp]
\begin{center}
\includegraphics[width=0.45\textwidth]{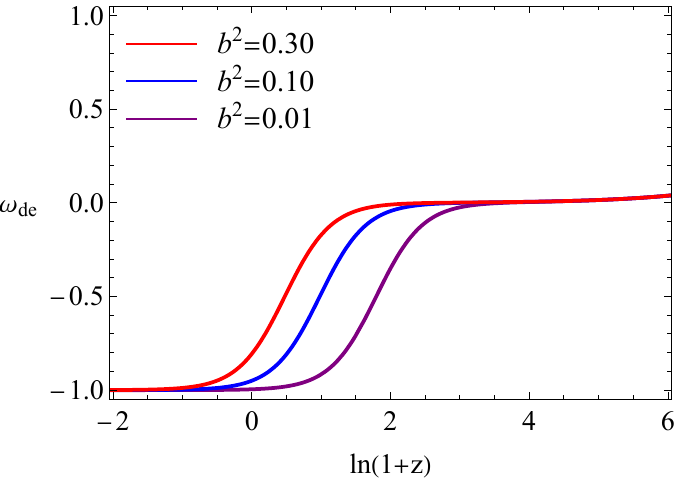}
\includegraphics[width=0.45\textwidth]{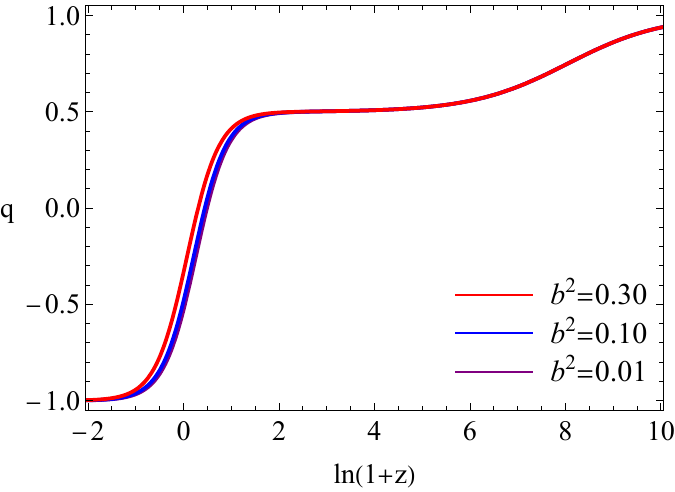}
\caption{\label{Fig5} Evolution curves of $\omega_{de}$ and $q$ versus redshift parameter $1+z$ for NHDEH.}
\end{center}
\end{figure*}

\begin{figure*}[htp]
\begin{center}
\includegraphics[width=0.45\textwidth]{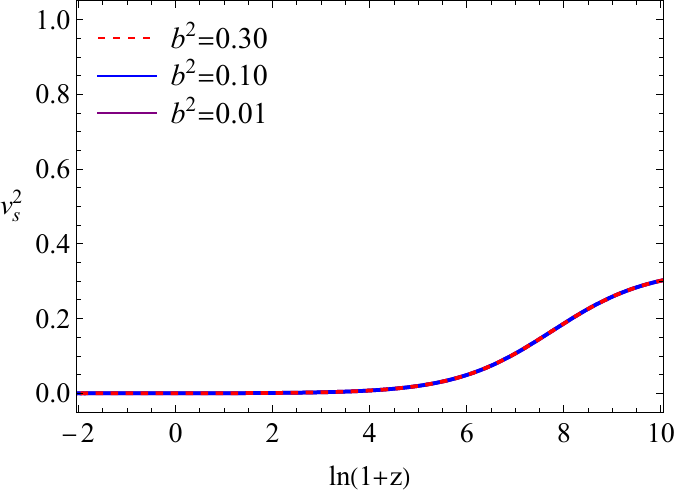}
\includegraphics[width=0.45\textwidth]{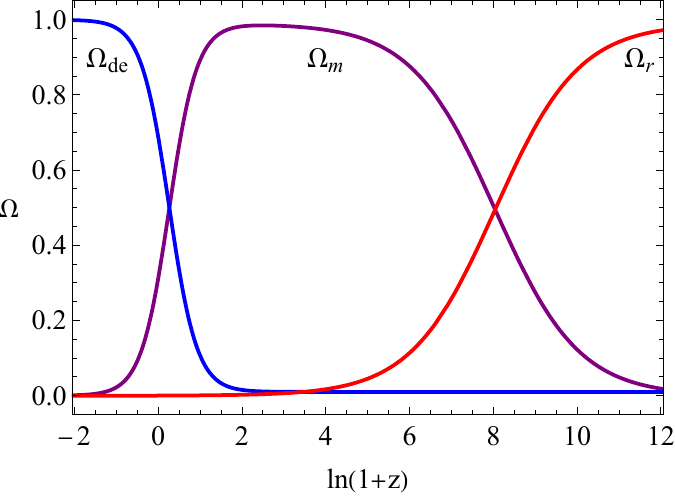}
\caption{\label{Fig6} Evolution curves of $v_{s}^{2}$ and density parameters $\Omega$ versus redshift parameter for NHDEH. The right panel is plotted for $b^{2}=0.01$.}
\end{center}
\end{figure*}

To analyze the evolution of the universe in NHDEH, we take $\Omega^{0}_{r}=0.0001$, $\Omega^{0}_{m}=0.3111$, $\Omega^{0}_{D}=0.6888$, and $H_{0}=67.66 km s^{-1}Mpc^{-1}$~\cite{Planck2020} as the initial conditions throughout this paper. Then, solving Eqs.~(\ref{Omm}) and~(\ref{Omde}) numerically, we obtain the evolution curves of $\Omega_{de}$, $\Omega_{m}$, $\omega_{de}$, and $q$, which are plotted in Figs.~(\ref{Fig4}) and ~(\ref{Fig5}). From Fig.~(\ref{Fig4}), we can see that $\Omega_{de}$ approaches $1$ and $\Omega_{m}$ approaches $0$ as the redshift decreases, and the universe can be dominated by NHDEH during its late-time evolution. However, in the relatively early stages of evolution, as $b^{2}$ increases, $\Omega_{de}$ increases while $\Omega_{m}$ decreases, indicating that a smaller $b^{2}$ is required to accurately describe the evolution of the universe in NHDEH. These results indicate that a small $b^{2}$ can be utilized to describe the evolution of the universe. The left panel of Fig.~(\ref{Fig5}) shows that the NHDEH behaves as quintessence, and $\omega_{de}$ approaches $-1$, behaving like a cosmological constant during the late-time evolution. The right panel of Fig.~(\ref{Fig5}) indicates that the late-time acceleration can be achieved, and a suitable range for the transition redshift($0.48\leq z_{t} <1$) is obtainable. In addition, the left panel of Fig.~(\ref{Fig6}) shows that this model is stable and the squared sound speed $v^{2}_{s}$ is independent of $b^{2}$.

Thus, according to the above results, since both $q$ and $\omega_{de}$ approach $-1$, the current acceleration can be realized, and NHDEH can behave as the cosmological constant. The universe will eventually evolve into an epoch described by the standard $\Lambda$CDM model, and the whole evolution of the universe can be described by this model, which is depicted in the right panel of Fig.~(\ref{Fig6}).

\subsection{Dynamical analysis}

In the previous subsection, we analyzed the evolution of the universe in NHDEH and found it to be stable against perturbations. In this subsection, we will analyze the dynamical behavior of NHDEH. In order to investigate the complete asymptotic behavior of NHDEH, we will adopt dynamical system analysis~\cite{Bahamonde2018, Wu2010, Dutta2017, Huang2019a, Huang2021, Wu2007, Wu2008}. The critical points can be obtained by solving the equations of the autonomous system
\beq
\Omega_{m}'=\Omega_{de}'=0.
\eeq
For the autonomous systems ~(\ref{Omm}) and ~(\ref{Omde}), we obtain three critical points as shown in Table~\ref{Tab1}. This table shows that the critical points $P_{1}$ and $P_{2}$ are determined by the value of $b^{2}$. When $b^{2}$ takes small values, point $P_{1}$ denotes the radiation dominated deceleration epoch, and NHDEH behaves as radiation; point $P_{2}$ represents the pressureless matter dominated deceleration epoch, and NHDEH behaves as pressureless matter; and point $P_{3}$ is an acceleration epoch dominated by NHDEH, and NHDEH behaves as the cosmological constant.

\begin{table*}
\caption{\label{Tab1} Critical points and stability conditions of NHDEH.}
\begin{center}
 \begin{tabular}{|c|c|c|c|c|c|c|c|}
  \hline
  \hline
  $Label$ & $Critical \ Points (\Omega_{m}, \Omega_{de})$ & $\Omega_{r}$ & $\omega_{de}$ & $q$ & $Eigenvalues$ & $Conditions$ & $Points$\\
  \hline
  $P_{1}$ & $(0,b^{2})$ & $1-b^{2}$ & $\frac{1}{3}$ & $1$ & $(4,1)$ & $Always$ & $Unstable \ point$\\
  \hline
  $P_{2}$ & $(1-b^{2},b^{2})$ & $0$ & $0$ & $\frac{1}{2}$ & $(3,-1)$ & $Always$ & $Saddle \ point$\\
  \hline
  $P_{3}$ & $(0,1)$ & $0$ & $-1$ & $-1$ & $(-4,-3)$ & $Always$ & $Stable \ point$\\
  \hline
  \hline
  \end{tabular}
\end{center}
\end{table*}

\begin{figure*}[htp]
\begin{center}
\includegraphics[width=0.45\textwidth]{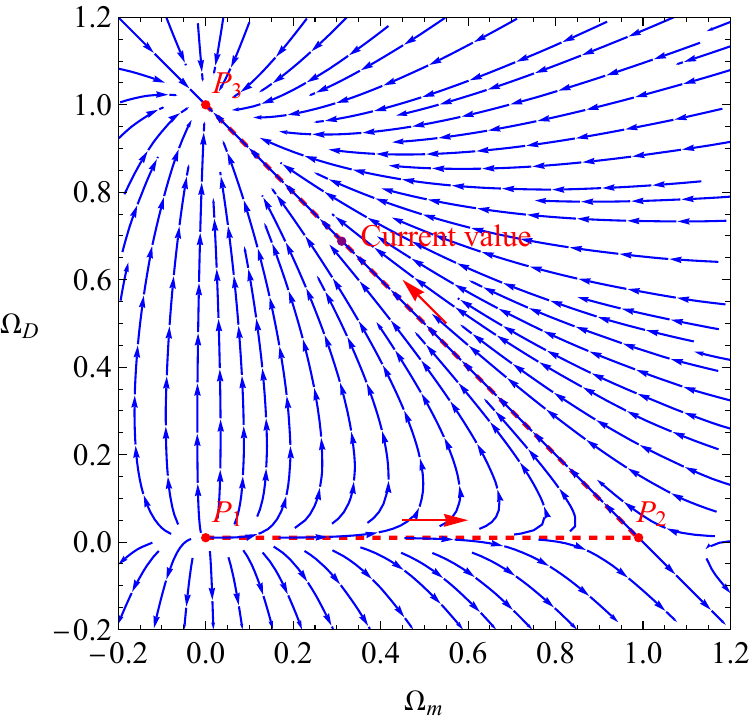}
\includegraphics[width=0.45\textwidth]{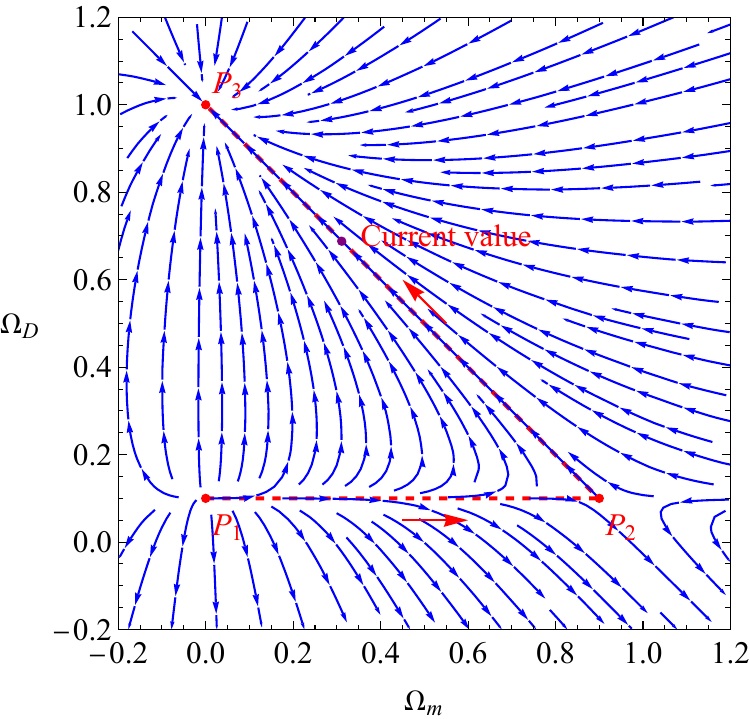}
\caption{\label{Fig7} Phase space trajectories and evolutionary curves of NHDEH. The left panel is plotted for $b^{2}=0.01$, and the right one is for $b^{2}=0.1$.}
\end{center}
\end{figure*}

Linearizing the equations of the autonomous system ~(\ref{Omm}) and ~(\ref{Omde}), we obtain the corresponding first order differential equations. The coefficient matrix of these equations determines the stability of the critical points, which are presented in Table~\ref{Tab1}.
For a critical point, if all eigenvalues are negative, it is a stable point representing an attractor; if all eigenvalues are positive, the corresponding point is unstable; if the eigenvalues have opposite signs, the critical point is a saddle point. From Table~\ref{Tab1}, we can see that point $P_{1}$ is unstable, $P_{2}$ is a saddle point, and $P_{3}$ is stable. According to the stability of these critical points, we can see that the universe evolves from the radiation dominated epoch $(P_{1})$ into the pressureless matter dominated epoch $(P_{2})$, and ultimately enters the NHDEH dominated late-time acceleration epoch $(P_{3})$. The behavior of attractor $P_{3}$ and the evolutionary trajectories are shown in Fig.~(\ref{Fig7}), where the left panel is plotted for $b^{2}=0.01$ and the right one is for $b^{2}=0.1$. Compared to the right panel of Fig.~(\ref{Fig7}), the left one is more suitable for describing the entire evolutionary history of the universe. Therefore, a smaller $b^{2}$ is required to describe the entire evolutionary history of the universe in NHDEH, and the universe will eventually evolve into an epoch characterized by the cosmological constant $\Lambda$.

\subsection{Statefinder analysis}

In the previous subsection, we have discussed the dynamical evolution of the universe in NHDEH using dynamical systems analysis and found that NHDEH can describe the entire evolutionary history of the universe for small values of $b^{2}$. Since an attractor behaving as the cosmological constant $\Lambda$ exists in NHDEH, distinguishing NHDEH from the standard $\Lambda$CDM model becomes an important issue. To address this issue, we will use the statefinder diagnostic pairs $\{r, s\}$, which were introduced by Sahni \textit{et al.}~\cite{Sahni2003}, to investigate whether NHDEH can be distinguished from the standard $\Lambda$CDM model.

\begin{figure*}[htp]
\begin{center}
\includegraphics[width=0.45\textwidth]{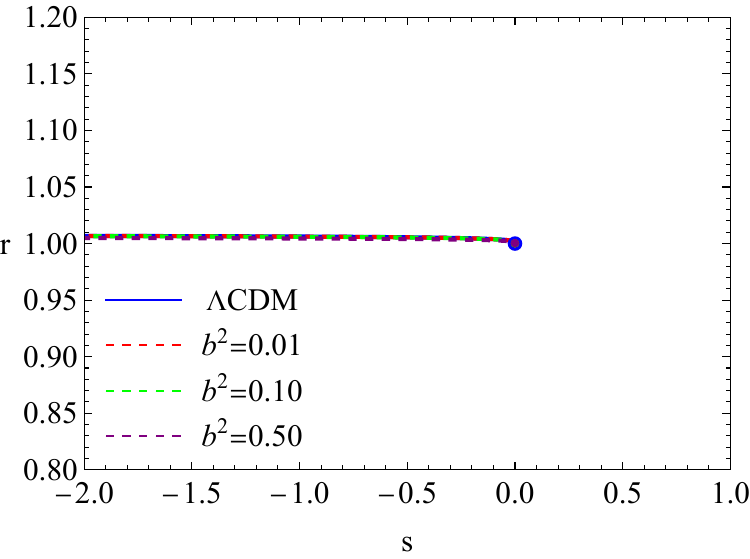}
\includegraphics[width=0.45\textwidth]{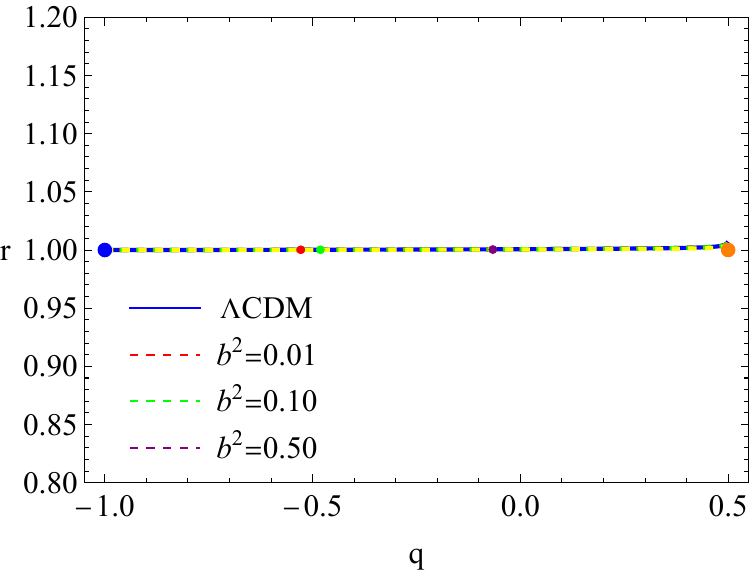}
\caption{\label{Fig8} Examples for the statefinder diagnostics $\{r, s\}$ and $\{r, q\}$ of NHDEH. The red, green, and purple dots represent the current values for $b^{2}=0.01$, $0.1$, and $0.5$.}
\end{center}
\end{figure*}

The statefinder parameters $r$ and $s$, which are geometrical diagnostics and depend only on $a$, are defined as~\cite{Sahni2003, Wu2005}
\beq
r=\frac{\dddot{a}}{a H^{3}}, \qquad s=\frac{r-1}{3(q-\frac{1}{2})}.
\eeq
Differentiating Eq.~(\ref{HH2}), the statefinder parameters $r$ and $s$ can be expressed in terms of $\Omega_{m}'$ and $\Omega_{de}'$. Thus, we can obtain the evolution curves of the universe in the $r-s$ parameter space by numerically solving Eqs.~(\ref{Omm}) and~(\ref{Omde}). In the left panel of Fig.~(\ref{Fig8}), we plot an example for the evolution curves of the statefinder diagnostic pair $\{r, s\}$. In this panel, the blue line represents the $\Lambda$CDM model, while the red, green, and purple dashed lines denote the evolutionary curves for $b^{2}=0.01$, $0.1$, and $0.5$; the blue dot marks the $\Lambda$CDM fixed point $(0,1)$, and the red, green, and purple dots represent the current values for $b^{2}=0.01$, $0.1$, and $0.5$. The results from the left panel of Fig.~(\ref{Fig8}) indicate that we cannot distinguish NHDEH from the standard $\Lambda$CDM model using the statefinder diagnostic pairs $\{r, s\}$, as all the evolutionary curves and points overlap. Then, we depict another statefinder diagnostic pairs $\{r, q\}$ in the right panel of Fig.~(\ref{Fig8}), in which the blue dot denotes the de Sitter expansion fixed point $(-1,1)$ in the future and the orange dot represents the standard cold dark matter fixed point $(0.5,1)$. This panel shows that all the curves overlap, starting from the standard cold dark matter fixed point $(0.5,1)$ and then evolving into the de Sitter expansion fixed point $(-1,1)$. Thus, using the statefinder diagnostic pairs $\{r, s\}$ and $\{r, q\}$, we cannot distinguish NHDEH from the standard $\Lambda$CDM model.

\subsection{Hubble diagram}

By considering the future event horizon as an IR cutoff, it is found that there exists a turning point in the Hubble diagram $H(z)$ in the holographic dark energy model~\cite{Colgain2021}. Subsequently, it is found that this turning point can be avoided in the Barrow holographic dark energy model under some specific circumstances~\cite{Huang2021}. In this subsection, we will discuss whether this turning point exists in NHDEH.

\begin{figure*}[htp]
\begin{center}
\includegraphics[width=0.6\textwidth]{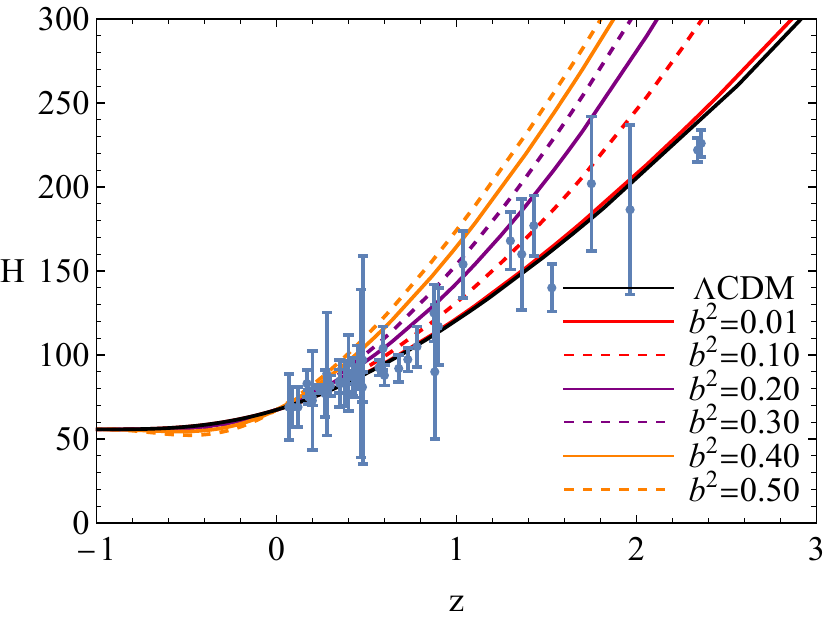}
\caption{\label{Fig9} Evolution curves of $H$.}
\end{center}
\end{figure*}

To achieve this goal, we have plotted the evolutionary curves of the Hubble parameter $H$ in Fig.~(\ref{Fig9}), where the error bars represent the observational Hubble parameter data~\cite{Akhlaghi2018, Cao2021}. This figure shows that the evolutionary curves of the Hubble parameter $H$ in NHDEH deviate from the curve of the standard $\Lambda$CDM model as $b^{2}$ increases, while they will overlap with the curve of the standard $\Lambda$CDM model when $b^{2}$ is very small; the turning point will disappear when $b^{2}$ takes small values. Thus, under the constraint $b^{2}=0.01$~\cite{Nakarachinda2022}, the turning point of the Hubble diagram in NHDEH is nonexistent.

\section{Conclusion}

Based on the entropy of anti-de Sitter black holes, a novel holographic dark energy model has been proposed. This model can realize the late-time accelerated expansion of the universe when the Hubble horizon is chosen as the IR cutoff. In this paper, by considering the Hubble horizon as the IR cutoff, we investigate holographic inflation and slow-roll inflation within this model. We find that holographic inflation is not supported by Planck results since $n_{s}$ is larger than $6.98$ for $r<0.06$. For slow-roll inflation with the chaotic potential $V_{0}\phi^{n}$, Planck results favor some specific cases, such as $n=\frac{1}{3}$ with $b^{2}=0.50$, and $n=\frac{1}{2}$ with $b^{2}=0.65$. Furthermore, for $n=\frac{1}{3}$, the observed value for $b^{2}=0.01$~\cite{Nakarachinda2022} is not excluded.

Subsequently, we investigate reheating in this model by calculating the reheating temperature and the number of reheating e-folds. Our results demonstrate that instantaneous reheating is excluded by the Planck results, as the corresponding spectral index $n_{s}$ exceeds the $2\sigma$ bound on $n_{s}$ from Planck data. For the cases $n=\frac{1}{3}$ and $n=\frac{1}{2}$ with the observed value $b^{2}=0.01$, the allowed ranges for the reheating temperature $T_{reh}$ and the number of reheating e-folds $N_{reh}$ are $10^{-2}GeV < T_{reh} < 10^{5}GeV$ and $9 < N_{reh} < 42$, respectively. These results are consistent with both theoretical predictions and observational constraints.

Finally, we analyze the evolution of the universe in this model and then apply dynamical analysis techniques to it. Our study indicates that the late-time acceleration of the universe can be achieved as $q$ and $\omega_{de}$ approach $-1$. This model is stable against perturbations, as the squared sound speed $v^{2}_{s}$ is positive. It can describe the entire evolutionary history of the universe when $b^{2}$ takes small values, with radiation, pressureless matter, and NHDEH successively dominate the evolution of the universe. Due to the presence of an attractor corresponding to the epoch described by the $\Lambda$CDM model, the universe will eventually evolve into an epoch characterized by the cosmological constant $\Lambda$. To distinguish this model from the standard $\Lambda$CDM model, we apply the statefinder analysis method. When using the statefinder diagnostic pairs $\{r, s\}$ and $\{r, q\}$ to discriminate this model from the standard $\Lambda$CDM model, we find that all the evolutionary curves and points overlap in the $r-s$ plane, and all the evolutionary curves overlap in the $r-q$ plane. Therefore, it is difficult to distinguish this model from the standard $\Lambda$CDM model using the statefinder diagnostic pairs. Then, we discuss the turning point of the Hubble diagram in this model. We find that the evolutionary curves of the Hubble parameter $H$ overlap with the curve of the standard $\Lambda$CDM model for sufficiently small values of $b^{2}$, and the turning point disappears when $b^{2}$ is small. In particular, for $b^{2}=0.01$, the turning point does not exist. These results indicate that when $b^{2}$ takes a small value, NHDEH cannot be distinguished from the standard $\Lambda$CDM model and can serve as an alternative model to the standard $\Lambda$CDM model.

\begin{acknowledgments}

This work was supported by the National Natural Science Foundation of China under Grants Nos.12405081, 12265019, 11865018, 12305056, the University Scientific Research Project of Anhui Province of China under Grants No. 2022AH051634.

\end{acknowledgments}

\end{document}